# Programmable phase selection between altermagnetic and non-centrosymmetric polymorphs of MnTe on InP via molecular beam epitaxy

An-Hsi Chen[1#], Parul R. Raghuvanshi[1#], Jacob Cook[1,2], Michael Chilcote[1], Jason Lapano[1], Alessandro R. Mazza[1], Qiangsheng Lu[1], Sangsoo Kim[1], Yueh-Chun Wu[1], T. Zac Ward[3], Benjamin Lawrie[1], Guang Bian[2], James Burns[3], Jonathan D. Poplawsky[3], Myung-Geun Han[4], Yimei Zhu[4], Lucas Lindsay[1], Hu Miao[1], Robert G. Moore[1], Gyula Eres[1], Valentino R. Cooper[1@], Matthew Brahlek[1*]

[1]Materials Science and Technology Division, Oak Ridge National Laboratory, Oak Ridge, TN, 37831, USA
[2]Department of Physics University of Missouri, Colombia, MO, 65201, USA
[3]Center for Nanophase Materials Sciences (CNMS), Oak Ridge National Laboratory, Oak Ridge, TN 37830, USA
[4]Condensed Matter Physics and Materials Science Department, Brookhaven National Laboratory, Upton, New York 11973, USA
[#]These authors contributed equally
Correspondence should be addressed to [@]coopervr@ornl.gov, [*]brahlekm@ornl.gov

**Abstract**: Phase selecting nearly degenerate crystalline polymorphs during epitaxial growth can be challenging yet is critical to targeting physical properties for specific applications. Here, we establish how phase selectivity of altermagnetic and non-centrosymmetric polymorphs of MnTe with high structural quality and phase purity can be programmed by subtle changes to the surface of lattice-matched InP substrates in molecular beam epitaxial (MBE) growth. Bulk altermagnetic MnTe is thermodynamically stable in the hexagonal NiAs-structure and is synthesized here on the (111)*A* surface (In-terminated) of InP, while the non-centrosymmetric, cubic ZnS-structure with wide band gap (> 3eV) is stabilized on the (111)*B* surface (P-terminated). Here we use electron microscopy, photoemission spectroscopy, and reflection high-energy electron diffraction, which together indicate that the phase selection is triggered at the interface and proceeds along the growing surface. First principles calculations suggest that interfacial termination and strain have a significant effect on the interfacial energy; stabilizing the NiAs polymorph on the In-terminated surface and the ZnS structure on the P-terminated surface. Selectively grown, high-quality films of MnTe polymorphs are key platforms that will enable our understanding of the novel properties of these materials, thereby facilitating their use in new applications ranging from spintronics to microelectronic devices.

This manuscript has been authored by UT-Battelle, LLC under Contract No. DE-AC05-00OR22725 with the U.S. Department of Energy. The United States Government retains and the publisher, by accepting the article for publication, acknowledges that the United States Government retains a non-exclusive, paid-up, irrevocable, world-wide license to publish or reproduce the published form of this manuscript, or allow others to do so, for United States Government purposes. The Department of Energy will provide public access to these results of federally sponsored research in accordance with the DOE Public Access Plan (http://energy.gov/downloads/doe-public-access-plan).

The manganese telluride family of magnetic semiconductors with bandgaps ranging from one to several electron volts and with magnetic transitions occurring above room temperature have long been of interest for applications in spintronics. The phase diagram shows that MnTe forms many polymorphs with the same stoichiometry with increasing temperature (see Ref. [1] and references therein). Figure 1(a) illustrates two of the most widely studied polymorphs of MnTe: the hexagonal nickel arsenide (NiAs; left) and the cubic zinc blende (ZnS; right) structures[1]. The NiAs structure of MnTe consists of alternate a-b-a-b stacking of octahedrally coordinated MnTe layers along the hexagonal <0001> direction and is the thermodynamically favored structure at ambient conditions. Whereas the cubic ZnS phase is composed of a-b-c-a-b-c stackings of tetrahedrally coordinated MnTe layers along the cubic <111> direction. Other polymorphs that have been reported include MnTe in the wurtzite structure, which also consists of tetrahedrally coordinated layers but with a-b-a-b stacking along the hexagonal <0001>. The NaCl structure that has octahedral coordination, like NiAs, but with a-b-c-a-b-c ordering along its cubic <111> direction. Here, we demonstrate that molecular beam epitaxy (MBE) offers a route to robust phase selectivity of the first two structures, NiAs and ZnS with varying electronic and magnetic behaviors, as it operates away from thermodynamic equilibrium.

The NiAs phase of MnTe is antiferromagnetic and has been reported to have a Néel temperature above room temperature and a band gap of 1.2-1.3 eV[2,3]. The antiferromagnetic ordering is A-type with ferromagnetic layers stacked antiferromagnetically along the $c$-axis[4]. Recently, NiAs-MnTe was predicted to be a prototypical altermagnetic material[5,6], a novel electronic phase where electronic bands of opposite spin are energetically split, and this spin-splitting *alternates* around the Fermi surface and closes along high symmetry directions[5–7]. This property is a consequence of the symmetries of both the crystalline and the magnetic lattices and the degree of spin-polarization driven by the atomic species and their orbital character. Early studies have shown that growing stochiometric NiAs-MnTe bulk crystals are difficult to grow due to a high density of defects that arise because it tends to favor being slightly Mn-rich[1]. These defects typically act as hole dopants in as-grown NiAs-MnTe producing bad metal behavior as well as being weakly ferromagnetic in many bulk and thin film samples [1,8–12].

NiAs-MnTe has room temperature lattice parameters of $a \approx 4.13$ Å and $c \approx 6.7$ Å[13] for which $a$ gives a close lattice match (0.5%) to the InP (111) lattice ($a \times \sqrt{2}/2 = 4.15$ Å, where $a$ is the InP cubic lattice parameter), as shown in Figure 1(c) left. In addition to reports of successful growth on InP[9–12], NiAs-MnTe has been grown on substrates including $SrF_2$[12], GaAs(111)[14], $Al_2O_3$(0001)[15,16] and Si(111)[17]. Clear challenges remain regarding the controlled synthesis of high quality, phase pure MnTe which is critical for confirmation and possible utilization of its altermagnetic properties. In addition to the novel aspect of altermagnetism, NiAs-MnTe has a strong coupling of the magnetism and crystalline structure, which has been explored for applications such as thermoelectricity and magnetoresistivity[9,12,13].

The ZnS structure is not the thermodynamically stable bulk phase, but has been grown as thin films[18–20]. Electronically ZnS-MnTe has a large band gap of ~3.1 eV[21]. First principles calculations were used to predict the properties of ZnS-MnTe as well as the phase stability through Cd-alloying[22,23]. The wide bandgap, located in the blue wavelength region, attracted interest in developing optical applications, such as barriers in quantum wells, $e.g.$, MnTe/CdTe/MnTe or MnTe/InSb/MnTe[24], and ultraviolet photodetectors[25]. There are limited reports of magnetic structure ZnS-MnTe, but it has been reported to exhibit antiferromagnetic ordering with a Néel temperature of ~65-70 K [26–28]. Interestingly, the ZnS structure is non-centrosymmetric, which allows for possible piezo- and ferro-electric functionalities. Coupling this with the magnetic ground state makes this an attractive platform for multiferroic studies.

Regarding growth of thin films of MnTe in the ZnS structure: it is cubic with lattice parameter of $a = 6.4$ Å[28], which translates to an effective in plane lattice parameter of 4.51 Å for the triangular (111) surface. MnTe stabilized in the ZnS structure was first reported using a CdTe substrate[21]. A later report showed epitaxial ZnS-MnTe directly grown on GaAs(001) which does not have a favorable lattice match (ZnS-MnTe(001)//GaAs(001) ($i.e.$, lattice mismatch ~12%)[29,30]. Of relevance to this work, ZnS-MnTe shows similarly large strain of -8.3% relative to the InP (111) surface. The ZnS-MnTe structure seems to be naturally stabilized when grown as a thin film under ultra-high-vacuum conditions, as there have been many reports on various semiconducting substrates such as GaAs[28], ZnTe substrates or epilayers[21], $SrTiO_3$(001)[20], mica[25], and $In_2Se_3$[16].

Here, we report a route to controllably stabilize phase-pure NiAs and ZnS polymorphs of MnTe on InP (111) substrates. The key finding is that the InP (111) surface termination triggers phase selectivity. Specifically, the In-terminated surface nucleates the NiAs phase, whereas the P-terminated surface nucleates the ZnS structure. The resulting films are found to be phase-pure and of high structural quality, as probed by both x-ray diffraction (XRD) and high-angle annular dark field scanning transmission electron microscopy (HAADF-STEM). Importantly, x-ray photoemission spectroscopy (XPS), HAADF-STEM, STEM electron energy loss spectroscopy (EELS), and atom-probe tomography (APT) indicate that the interfaces are of high crystalline quality with minimal chemical intermixing. First principles calculations show that bulk NiAs-MnTe has the lowest energy (as expected) yet is close to that of bulk ZnS-MnTe. However, on the InP(111) surface, the interfacial energy favors NiAs-MnTe on In-terminated InP and ZnS-MnTe on P-terminated InP. In summary, In-terminated surfaces promote the initial growth of NiAs-MnTe which continues as the film grows, whereas P-termination favors ZnS-MnTe growth. This work not only demonstrates an important result concerning phase selectivity by MBE growth but also proves vital for advancing the understanding and utilization of MnTe polymorphs with novel properties for possible future applications.

*Film synthesis and characterization* – The MnTe samples were grown in a home-built, MBE system with base pressure $<1\times10^{-9}$ Torr and equipped with *in situ* reflection high energy electron diffraction (RHEED) operated at 10 kV, as reported previously[11]. The substrates used were semiconducting Fe-doped InP(111)*A* (In-terminated) and InP(111)*B* (P-terminated) and the preparation of the surfaces prior to MBE growth is a crucial step that will be described later. Prior to growth, the sources were calibrated using a quartz crystal microbalance with a Mn flux of $\sim 2\times10^{13}$ cm$^{-2}$s$^{-1}$ and a Te flux $>7\times10^{13}$ cm$^{-2}$s$^{-1}$ at the sample surface. XRD was performed on a 4-circle Panalytical X'pert diffractometer using Cu-K$_{\alpha 1}$ radiation. Electron microscopy was performed using a JEOL ARM 200CF with a cold field emission gun and spherical aberration correctors operated at 200 kV.

The native oxide removal from the InP substrates prior to MBE growth is a critical step for growing high crystalline quality MnTe epitaxial thin films. A standard step in surface preparation for MBE growth of III-V semiconductors is an *ex-situ* chemical oxide strip combined with *in-situ* exposure to an overpressure group-V flux, like As, P, or Sb, at high temperatures. This high-temperature process then leaves a pristine surface for III-V semiconductor film growth. Similarly, for chalcogenide thin films on III-V semiconductor substrates *ex-situ* chemical oxide strip combined with *in-situ* thermal annealing in an overpressure of chalcogenide flux, such as Se or Te was used to prepare the surface for epitaxial growth[10,31,32]. However, we have found that aggressive chemical etching and the high-temperature anneal typically leaves pits or In blobs on the surface. Consequently, the following modified recipe for the removal of both organics and the oxide layer from the InP surface was developed and reported in Ref.[11] and follows a similar recipe for InP (001) surface[33]: Before loading the InP(111) substrates into the vacuum chamber, both surface terminations were exposed to UV-generated ozone for 10 minutes to remove organics. The oxide was then stripped by dipping the substrates into 5% HCl for 30 seconds for In-terminated or 5 seconds for P-terminated. The difference in time for HCl etching was found to be crucial for preserving smooth surfaces for both terminations[34]. Prolonged HCl etching (30 seconds) was found to roughen the P-terminated face as shown by RHEED in Figures S1-2. The In-terminated surface was less sensitive to the HCl etching time as the RHEED patterns for 5- and 30-seconds appear similar.

To assess the effectiveness of the oxide removal step and determine the chemical state of the different surface terminations we employed XPS to examine the P2$p$, In3$d$, and Te3$d$ states for both InP terminations (Figure 2a for In-terminated and Figure 2b for P-terminated) at each step of the surface preparation. The spectra for the as-received samples are the blue curves shown at the bottom of each panel. The P2$p$ shows double peaks corresponding to P2$p_{1/2}$ and P2$p_{3/2}$ with peak offset 0.87 eV, as well as In3$d$ exhibiting a characteristic In3$d_{3/2}$ and In3$d_{5/2}$ with peak offset of 7.55 eV, coming from the intrinsic spin-orbit splitting. Both elements show clear oxide peaks, P$_{ox}$ and In$_{ox}$, as highlighted with arrows and labels. We note only a slight difference in the intensity of the oxide peaks which are slightly larger for In-terminated than for P-terminated. The green colored curves in the middle row show that the oxide was completely removed by HCl etching. The key difference is that the spectra are slightly shifted to higher binding energies compared to the as received surface. The final data shown in Figure 2 (red curves, ii), where the samples were heated to the annealing temperature 250 °C, exposed to Te, and then cooled, reflect the surface prior to growth. Both samples exhibit identical Te$3d_{3/2}$ and Te$3d_{5/2}$ peaks after exposure to Te, which shows that a terminating layer of Te is present prior to growth. XPS spectra of the P and In peaks were found to shift slightly after Te exposure, which implies the surfaces have a different chemical environment due to Te termination[35]. Note that all the XPS spectra were aligned to the C1$s$ peak (Figure S3 and S4) to compensate for surface charging effects. These data demonstrate that the surfaces of both terminations are oxygen-free and the surfaces are terminated with Te prior to growth. The surfaces are found to be quite similar chemically with the only discernible difference being a slight chemical shift between the In and P states. In-situ RHEED was used to characterize the crystalline quality and morphology of the surfaces, and the images for the In and P terminated surfaces following the oxide-strip procedure but before annealing and exposure to Te are shown in Figure 3(b) column i) taken at room temperature. The images show diffraction peaks that are sharp and localized to the first Ewald circle as well as clear Kikuchi lines, which are both consistent with highly crystalline surfaces that are flat and two-dimensional.

Following the surface treatment of the InP, MnTe was grown by MBE according to the procedure that is schematically shown in Figure 3(a). The samples were heated to 250 °C then exposed to Te flux ($>7\times10^{13}$ cm$^{-2}$s$^{-1}$) for 15 mins. This temperature was well-above the desorption temperature for Te, thus there was no accumulation. As shown in Figure 3(b) column ii), RHEED images of the surface were nominally unchanged after Te exposure. After Te exposure, 2 monolayers of MnTe were deposited, which served as a homoepitaxial nucleation layer and was found to be critical for achieving nearly atomically flat surfaces with minimum roughness. Following this layer, the samples were heated to 325 °C in a Te flux where the second layer growth was performed to the desired thickness. 325 °C is the highest temperature that the surface of InP could be maintained confidently. For example, around 375 °C, some samples exhibited subtle streaks on the surface from excess In due to P desorption. For temperatures higher than 400 °C, there was noticeable formation of In clusters on the surface. Figure 3(b) column v) shows RHEED images of two MnTe thin film at room temperature. Both images consist of well-formed peaks that are streaky along the vertical, which indicates high-quality, flat (2D-like), crystalline surfaces; interestingly, the RHEED image of the NiAs-MnTe phase was slightly more diffuse compared to the ZnS-MnTe phase. This diffuseness suggests slightly more crystalline disorder despite the fact that NiAs is the stable bulk phase and possibly relates to the thermodynamically expected excess Mn.

MnTe with the NiAs and ZnS structures can be clearly distinguished by their in-plane and out-of-plane lattice parameters. This can be done by measuring the in-plane interplanar spacing with *in-situ* RHEED as well as both the out-of-plane and in-plane lattice parameters using XRD. First, comparison of the horizontal spacing of the RHEED streaks of MnTe relative to InP enables determining the in-plane spacing in real time during growth[16]. Figure 3(b) column v) shows that the in-plane lattice parameter of MnTe for NiAs-MnTe [11-20] is ~4.20±0.08 Å and ZnS-MnTe [110] is a bit larger at ~4.40±0.08 Å. This change in lattice parameter was observed starting at the first layer, indicating that the phase is determined at the interface with the substrate surface. Figure 3(c) shows XRD $2\theta$-$\theta$ scans for both the NiAs and ZnS structures along the InP(111) reflection, which is marked by the asterisks (see supplemental Figure S6 for scans with $2\theta$ over wider ranges). Both scans show reflections from the films, marked by vertical dashed lines. About the small periodic peaks surrounding the main reflections pendellösung fringes can be seen, a signature of atomically flat and parallel surfaces of the films in agreement with the RHEED analysis[36]. Furthermore, the two structures have quite different out-of-plane lattice parameters: the (111) reflection is at $2\theta$ ~ 24.2° for the sample grown on the P-terminated surface, consistent with the known lattice parameter of ZnS-MnTe, $a = 6.4$ Å[28]; for the sample grown on the In-terminated surface, the (0002) reflection occurs at $2\theta$ ~ 26.7°, consistent with the known lattice parameter of NiAs-MnTe, $c \approx 6.7$ Å[13]. Note that no other MnTe polymorphs were observed. For example, reflections from the NaCl phase would appear at $2\theta$ ~ 26.3° for the (111) reflection and the wurtzite phase would be close to $2\theta$ ~ 25.6° for the (0001) reflection. These data show that our oxide stripping and 2-step MBE growth recipe results in high quality single phase MnTe thin films of both NiAs and ZnS varieties depending on surface termination.

*Interfacial characterization* – To understand the interface between InP and the various MnTe phases, cross-sectional HAADF-STEM and APT that give insight into the local structural and chemical properties. Figure 4(a-b) shows HAADF-STEM for both NiAs-MnTe and ZnS-MnTe, while Figure 5 shows the APT measurement and analysis. First, MnTe (as previously reported[11]), in both the NiAs and the ZnS phases, exhibits clear structure and well defined interfaces indicative of high-quality crystalline structures, consistent with both XRD and RHEED measurements. Moreover, the HAADF-STEM images for the NiAs and ZnS structures show several notable features that are vital to understanding why each polymorph is stabilized, as discussed later. Second, the ZnS and NiAs structures are distinguishable by clear a-b-a-b ordering for the NiAs-MnTe phase compared to the ZnS-MnTe a-b-c-a-b-c ordering. The overall intensity of the NiAs-MnTe film is brighter than the substrate, which is consistent with Z-contrast from the larger nuclear charge density for MnTe relative to InP. For the MnTe in the ZnS structure, the Z-contrast is nominally the same as the substrate, which is consistent with the lower density compared to the NiAs phase. However, the resolution near the interface is sufficient to track atomic ordering. For NiAs the Z-contrast indicates that the substrate terminates with Te, which can be seen by the extracted line profile in the right panel of Figure 4(a). In contrast, the ZnS sample shows a unique multilayer structure across the interface which is highlighted by a sky-blue region in Figure 4(b). The interface appears to be a double layer of heavy elements, either Te or In, and the dark region in between could be either unfilled space or lighter P/Mn. STEM-EELS (Figure 4(c-d)) measurements show that this interface is likely composed of a lower layer of atoms that is In with possibly a small amount of Te mixed in and the upper layer is Te with a small amount of In mixed in. This potentially arises from the higher propensity of P-terminated surfaces to have free In on the surface[34,37] combined with the absence of direct Te-P bonding. Interestingly, the relative ratios of Te to P in the second layer and the initial layer of MnTe indicate that there is direct Te-P bonding, consistent with the XPS analysis.

APT utilizes an atomically sharp tip created out of the material of interest. A large electric field applied to the tip causes individual atoms to desorb and accelerate in the field and impinge upon a detector. The resulting detector positions can be uniquely traced back to the atomic positions in the sample with nanometer scale accuracy as well as the determination of the mass of the atom[38,39]. This gives an accurate picture of the atomic species and their locations across atomic scale interfaces. Figure 5(a) shows the atomically resolved positions (laterally, $x$ and across the interface, $z$) of both MnTe phases. The lower purple section is InP and the upper yellow section is MnTe. Both samples have a clear interface and no clear interdiffusion of In or P into MnTe layers. The chemical structure across the interfaces can be more clearly seen in the curves of Figure 5(b), where the compositions were integrated across the lateral dimension and normalized. These curves show that the interface is nearly abrupt, 1-2 nm, and there is minimal In or P uptake into the film, which eliminates possible chemical mixing as a driver of the stabilization of the ZnS phase, as was observed with Ga doping in MnAs grown on GaAs[40,41]. Moreover, the curves for both phases of MnTe are identical and directly overlap confirming that there is no chemical difference among the two phases.

*Theoretical modeling* – To gain deeper atomistic view of the stabilization of a NiAs and ZnS polymorphs of MnTe with varying InP terminations we modeled the interfacial structures and energies using density functional theory (DFT) as implemented in the Vienna Ab Initio Simulation Package (VASP)[42,43]. Details regarding DFT thresholds and parameters can be found in the Supplementary Information (SI). Here we provide structural details of the MnTe/InP heterostructures examined. The systems were modeled as slabs of MnTe and InP each consisting of 4 formula units perpendicular to their interface and $3 \times 3$ formula units parallel (see Figure 6). To avoid interactions between interfaces in the periodic VASP DFT simulations, a 40 Å vacuum separation was included in the direction perpendicular to the interface. Calculations included a Hubbard U correction for Mn $3d$ states ($U$=5 eV) and were spin-polarized with initial A-type antiferromagnetic order for Mn atoms (4.5 µB; zero moment for other atoms), see SI for more details. The in-plane lattice parameters for both NiAs and ZnS MnTe polymorphs were lattice matched to InP with lattice parameter described above. Calculations were initiated with the MnTe and InP slabs close enough for covalent bonding and then internal atomic coordinates were relaxed, while keeping the atoms in the bottom two atomic layers of the InP fixed to simulate bulk-like substrate behavior. To avoid unphysical charge accumulation and depletion and limits of formation from artificial dipoles, for both NiAs and ZnS MnTe polymorphs only Mn-termination (Te-termination) was considered with P-terminated InP (In-terminated InP).

Several important insights emerge from the DFT calculations of the various MnTe/InP interfaces. First, NiAs-MnTe consistently exhibits lower total energy than ZnS-MnTe on the InP substrate, regardless of termination, in agreement with the fact that NiAs-MnTe is the thermodynamic ground state. The black text below each panel in Figure 6 gives the total energy of each configuration. To assess the impact of surface termination on interface stability, we computed the interface energy density, $E_{int}$, for each MnTe/InP configuration using the following expression:

$$E_{int} = (E - E_{MnTe} - E_{InP})/A, \quad \text{(eq. 1)}$$

where E is the total energy of the combined MnTe/InP slab, $E_{MnTe}$ and $E_{InP}$ are the total energies of the isolated MnTe and InP slabs (fixed to their MnTe/InP slab relaxed geometries), respectively, and *A* is the interfacial area. This formulation isolates the energetic contribution from the interface itself (values shown in red in Figure 6). For the In-terminated InP substrate, the interface energy density was found to be 1 meV/Å$^2$ lower for NiAs-MnTe (-11 meV/Å$^2$) compared to ZnS-MnTe (-10 meV/Å$^2$), suggesting that the NiAs polymorph forms a marginally more favorable interface with the In-terminated surface. In contrast, for the P-terminated substrate, the ZnS-MnTe (-37 meV/Å$^2$) exhibited a significantly lower interface energy density of 5 meV/Å$^2$ than the NiAs-MnTe (-32 meV/Å$^2$), indicating higher propensity for the interfacial stability for ZnS in this configuration. These calculations indicate that although NiAs-MnTe has a net lower energy compared to ZnS-MnTe, the interface may play a dominate role for selecting a specific polymorph.

*Discussion and Conclusion* – Our study uncovers a striking and controllable phase selection mechanism for two functional polymorphs of MnTe during MBE growth, governed by the surface termination of the InP(111) substrate. Specifically, In-terminated surfaces stabilize the NiAs structure, while P-terminated surfaces favor the ZnS structure. We present compelling evidence that this phase selection is triggered at the interface—at the interface the structure is locked and then continues during subsequent film growth. This interfacial control, along with the high crystalline and chemical quality of the resulting films, is validated through XRD, XPS, RHEED, APT and HAADF-STEM analyses. Complementary DFT calculations confirm that while NiAs-MnTe is the thermodynamic ground state in bulk, ZnS-MnTe is close in energy. Importantly, the calculated interfacial energies reveal that NiAs-MnTe forms a more favorable interface with In-terminated InP, whereas ZnS-MnTe is preferred on P-terminated surfaces. Once nucleated, each phase exhibits self-propagating growth, although rare ZnS intergrowths are observed in NiAs films (~1 in 10), while ZnS films remain consistently phase-pure. These findings highlight how subtle differences in interfacial energetics can decisively influence polymorph selection during epitaxial growth.

Although the exact mechanism remains unresolved, there are several critical factors that contribute to this novel phase selectivity. Pauling's rules for crystals structures rely on the ratio of the cation and anion ionic radii ($r_i$) to predict crystal structures[44]. Applying these to MnTe ($Mn^{2+}$ with coordination of 4 ($r_i$ = 0.8 Å) and 6 ($r_i$ = 0.97 Å) and $Te^{2-}$ ($r_i$ = 2.06 Å))[44–46] yields a structure borderline between tetrahedral (ZnS) and octahedral (NiAs), cation to anion ratio ~0.414, which corroborates the DFT results. Since MnTe is close to the borderline between these coordinations implies that slight shifts in interfacial chemistry could play an important role through the effective Mn valence, and, thus, the ionic radius. Considering this in terms of the HAADF-STEM and STEM-EELS results, Pauling's rules may suggest a mechanism for the phase selectivity, which may be tied to slight differences in chemistry. Finally, MBE's lower growth temperature introduces kinetic limitations that favor ZnS-MnTe, possibly because the system is unable to reach the global energetic minimum required to stabilize NiAs-MnTe. Understanding how these subtle structural and chemical differences seed distinct polymorphs will be central to gaining full control over phase selectivity in MnTe and related systems.

To conclude, we have demonstrated that surface termination of InP(111) selectively stabilizes MnTe polymorphs (NiAs or ZnS), which result in films with high structural quality and atomically sharp interfaces. These results are both surprising as well as critical for future applications and scientific studies. For example, there has been a large push to find and understand altermagnetic materials and NiAs-MnTe is among the leading candidates. However, in bulk crystals the phase diagram shows that NiAs-MnTe is inevitably non-stoichiometric (either Mn-rich defects or $MnTe_2$ intergrowths). This work highlights the unique power of MBE growth to access conditions that differ significantly from bulk thermodynamics, thus providing hope to synthesize higher quality materials where intrinsic properties can be studied.

Furthermore, multifunctional magnetic-semiconducting and multiferroic materials have long been sought for materials-physics, where magnetism can be voltage controlled. The fields, respectively, have long focused on III-V semiconductors and transition metal oxides. However, ZnS MnTe seems to be the phase favored by MBE growth and integrable with standard semiconductor platforms. Thus, ZnS-MnTe offers a promising platform for exploring the coupling of magnetism and charge, though many questions remain regarding the functionality of its antiferromagnetic ground state and its overall usefulness. Moreover, the wide band gap and the non-centrosymmetric structure makes ZnS-MnTe an exciting candidate for possible multiferroics. Although the III-V ZnS structures are only weakly piezoelectric, understanding the coupling of antiferromagnetism and ferroic properties as well as possibly proximity enhancements remains exciting[47]. The ability to deterministically synthesize pristine films of either MnTe polymorph establishes a foundation for comprehensive investigation of their intrinsic properties and accelerates their development for emerging technologies.

**Data Availability**
The data supporting this study's findings are available from the corresponding author upon reasonable request.

**Supplementary Information**
Supplementary Information provides the experimental and theoretical methods, and addition RHEED, XPS, Raman and XRD data.

**Acknowledgments**

*This work is dedicated to our colleague and friend Jon Poplawsky who tragically passed away during the preparation of this work.*

This work was supported by the U. S. Department of Energy (DOE), Office of Science, Basic Energy Sciences (BES), Materials Sciences and Engineering Division (growth, structure, transport, and density functional calculations), and the National Quantum Information Science Research Centers, Quantum Science Center (spectroscopy). The work at the Brookhaven National Laboratory was supported by the U.S. Department of Energy (DOE), Basic Energy Sciences, Materials Science and Engineering Division, under Contract DESC0012704. The cross-sectional TEM samples were prepared by focused ion beam at the Center for Functional Nanomaterials, Brookhaven National Laboratory. G.B. was supported by the U.S. Department of Energy, Office of Science, Office of Basic Energy Sciences, Division of Materials Science and Engineering, under Grant No. DE-SC0024294.

**Figure 1:**

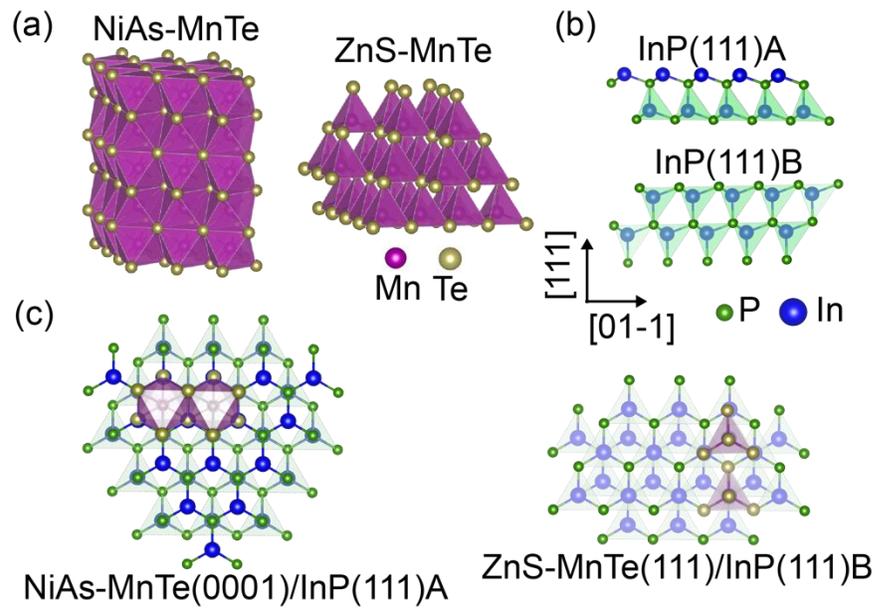

**Fig. 1** Illustration of select crystalline phases of MnTe in relation to the InP(111) surface. (a) NiAs- and ZnS-phase of MnTe from left to right. (b) The schematics of InP(111) with different terminations, In (top, InP(111)*A*) and P (bottom, InP(111)*B*) (c) The epitaxial relationship of MnTe on the InP(111) substrate, (left) NiAs-MnTe(0001)/InP(111)*A* (In-terminated) and (right) ZnS-MnTe(111)/InP(111)*B* (P-terminated).

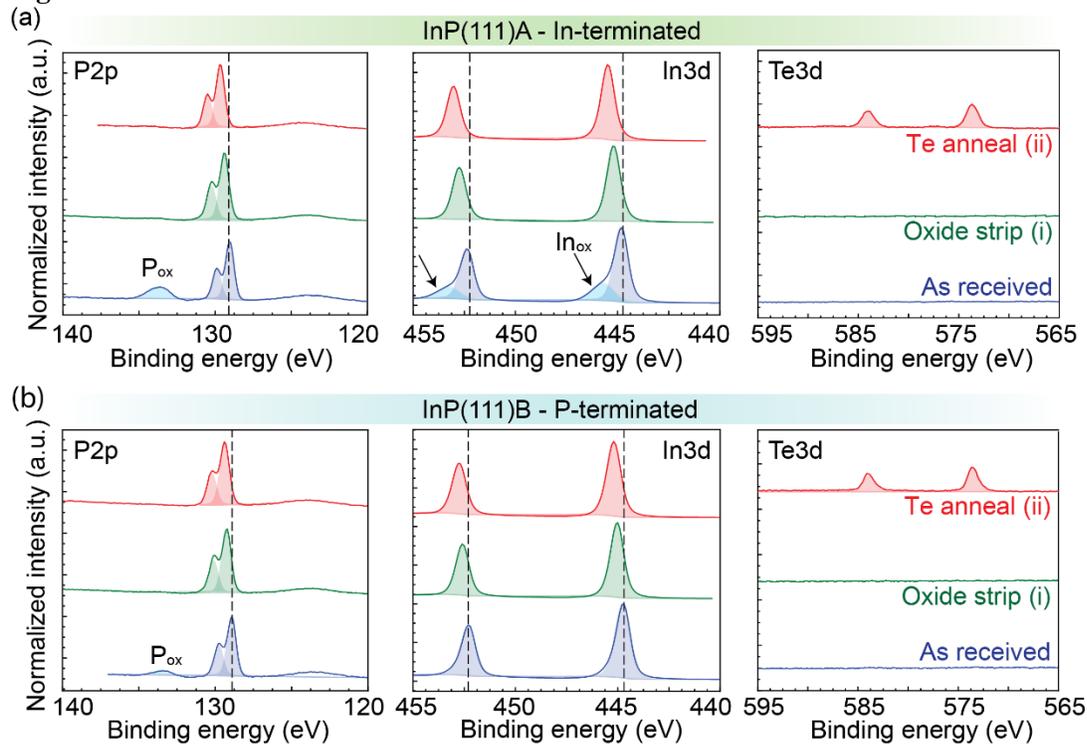

**Fig. 2.** XPS P2p, In3d and Te3d spectra from (a) InP(111)A (In-terminated) and (b) InP(111)B (P-terminated) after each step of the sample preparation. Spectra for as received, oxide strip (i) and oxide strip with a Te anneal at the growth temperature (ii) are indicated (i and ii corresponding to the labels in in Figure 3).

**Figure 3**

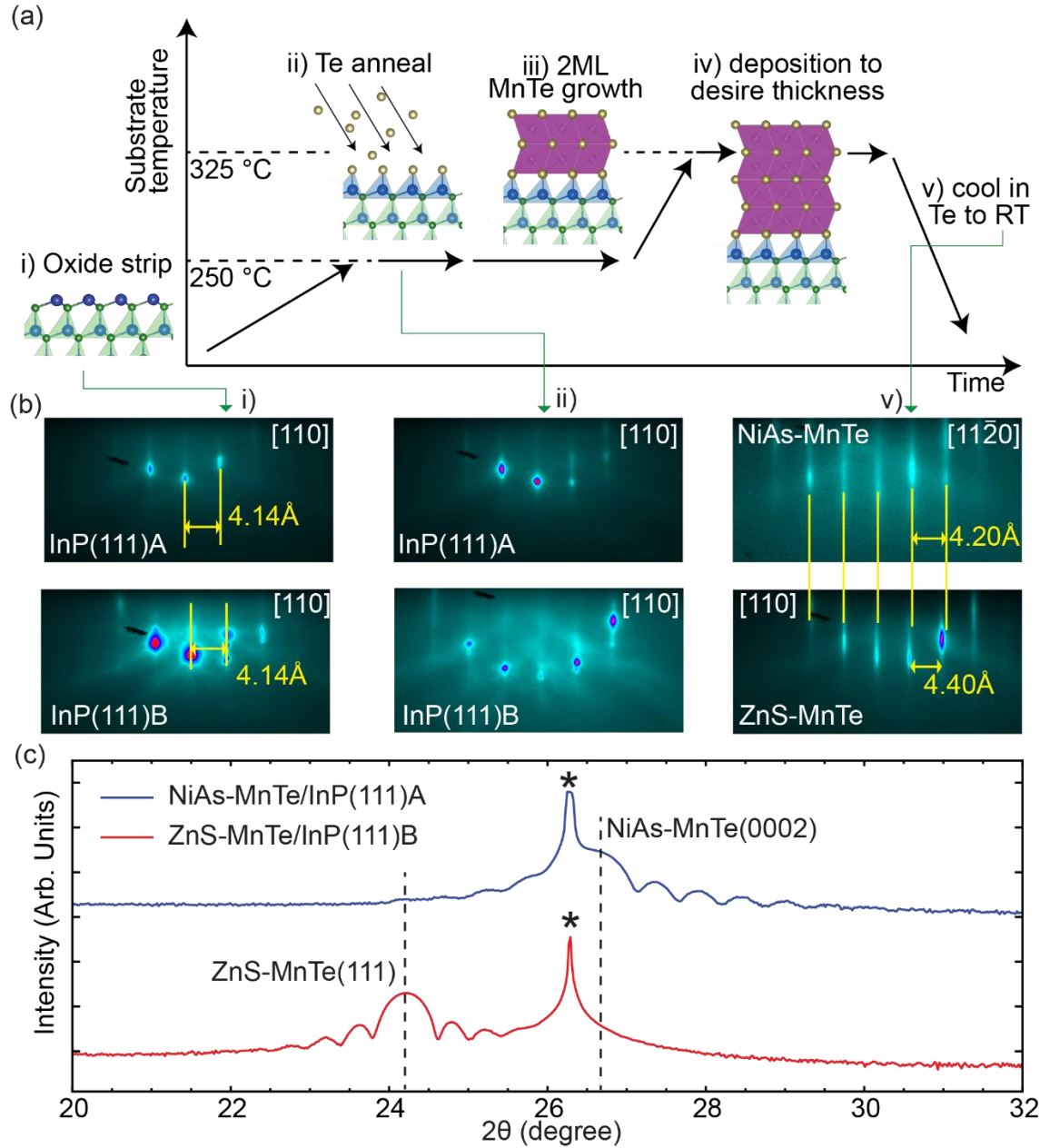

**Fig. 3** (a) Schematic showing of the optimized growth steps for high crystalline quality MnTe. (b) In-situ RHEED patterns are recorded during the growth of NiAs-MnTe/InP(111)A (In-terminated, first row) and ZnS-MnTe/InP(111)B (P-terminated, second row). Step i) represents the InP surface after HCl treatment, step ii) represents the InP surface after annealing in Te flux at 250°C, and step v) represents the MnTe surface at room temperature. (c) 2θ-θ measurement on NiAs-MnTe/InP(111)A (In-terminated) in blue, and ZnS-MnTe/InP(111)B (P-terminated) in red. InP(111) diffraction is indicated by the asterisk (see supplementary information for full-range 2θ-θ measurements).

**Figure 4**

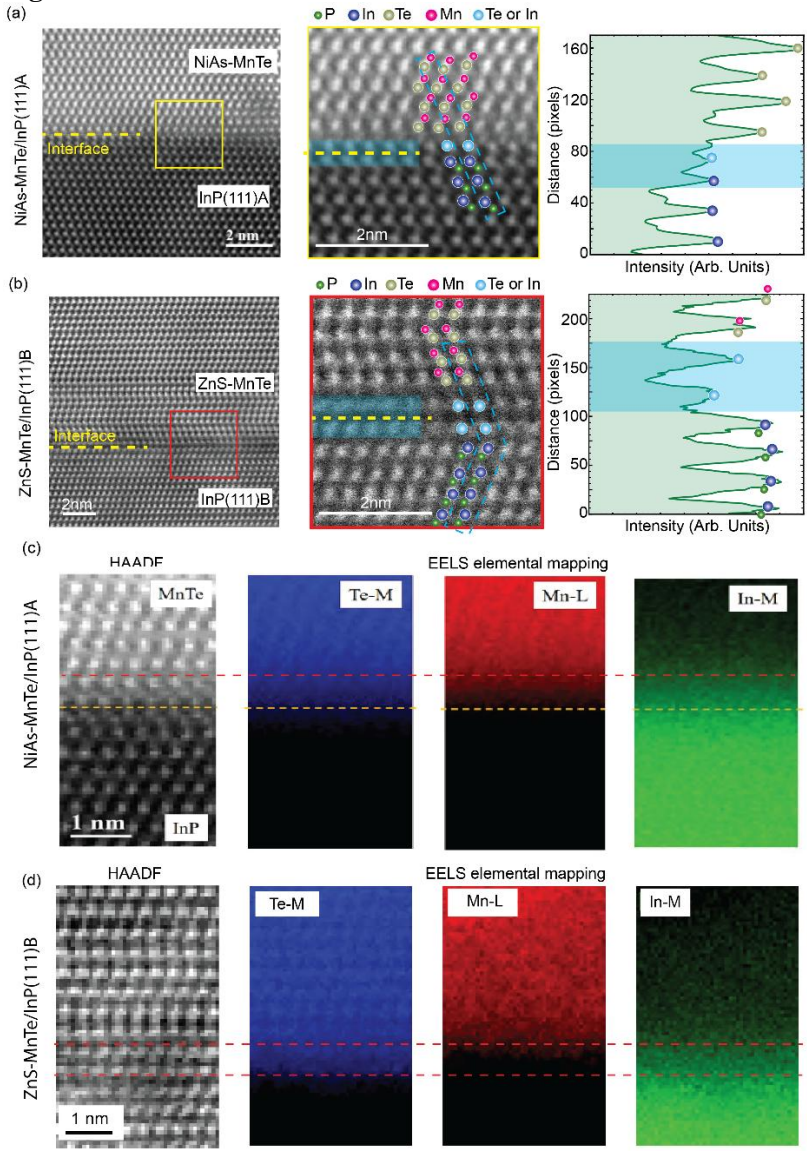

**Fig. 4** Interfacial and elemental study on NiAs-MnTe/InP(111)A (In-terminated face) and ZnS-MnTe/InP(111)B (P-terminated face). (a) Left: cross section of NiAs-MnTe/InP(111)A. Middle: the zoomed image at the interface highlighted in yellow solid square from the left image. Right: The extracted line profile from the middle image highlighted with green dashed region. (b) Left: the cross section of ZnS-MnTe/InP(111)B. Middle: the zoomed view highlighted in red solid square from the left image. Right: The extracted line profile from the middle image highlighted with green dashed region. The light blue circles represent either Te or In, as it is hard to determine from the intensity of STEM image. The interface is indicated by a yellow horizontal line. The light blue shaded region in (a) and (b) highlights the interfaces of MnTe/InP(111). The HAADF and EELS for (c) NiAs-MnTe/InP(111)A and (d) ZnS-MnTe/InP(111)B.

**Figure 5**

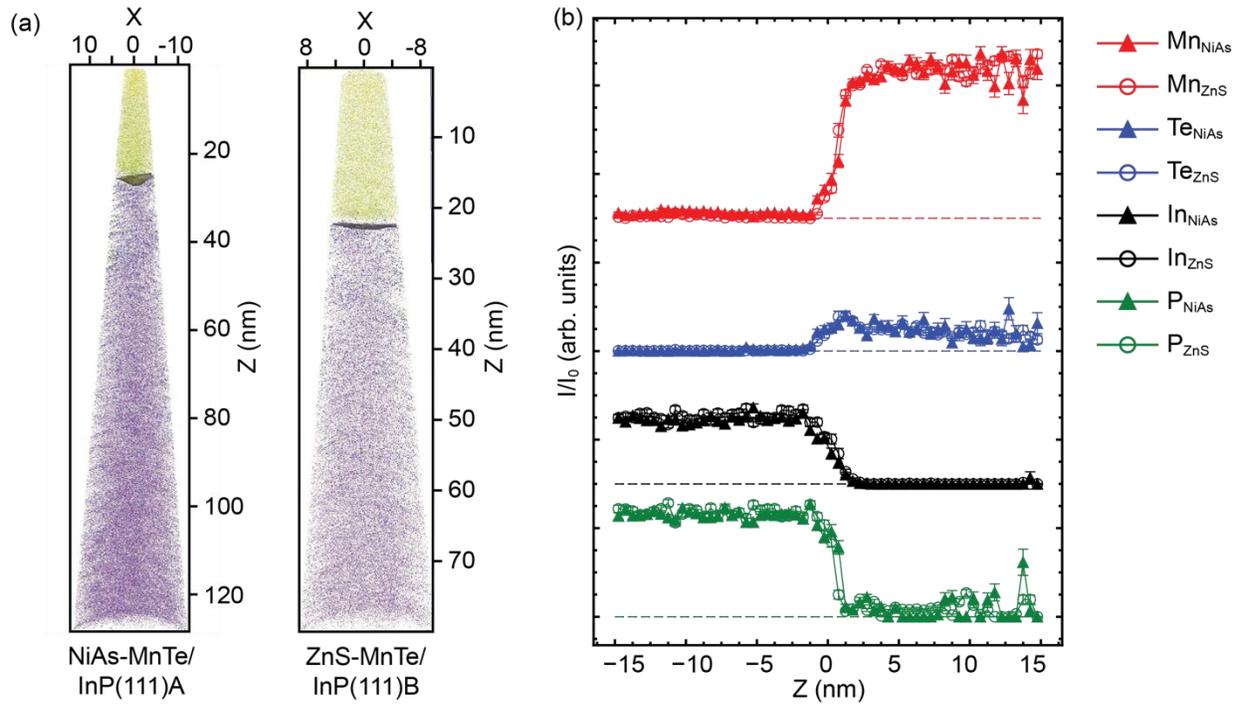

**Fig. 5** APT mapping of both phases of MnTe. (a) The APT tips for both phases of MnTe, and (b) normalized atomic percentage $I/I_0$ across the APT tips which shows the abruptness of MnTe-InP interface. From top to bottom, the curves represent: Mn, Te, In and P. The filled symbols are NiAs-MnTe and the open symbols are ZnS-MnTe.

**Figure 6**

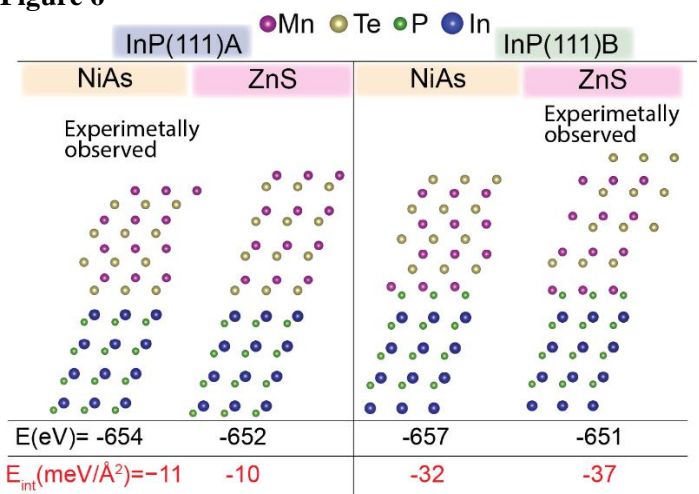

**Fig. 6** The total energy calculations of the MnTe-InP slabs for the NiAs and ZnS polymorphs and with different In-termination (InP(111)*A*, left) and P-termination (InP(111)*B*, right). Black values (top row) indicate total energy in eV and the interface energies, $E_{int}$, are red values (bottom row) in eV/Å$^2$.